%% file: ms.tex
\documentclass[10pt,conference,english,doublecolumn]{IEEEtran}
\usepackage[centertags]{amsmath}
\usepackage{amsfonts}
\usepackage{amssymb}
\usepackage{epsfig}
\usepackage{amsmath,  amssymb}
\usepackage{graphicx}
\usepackage[centertags]{amsmath}
\usepackage{clrscode}
\usepackage{cite}
\usepackage{fullpage}
\usepackage{amsmath,amssymb,mathrsfs,pseudocode}
\usepackage{color}
\usepackage[normalem]{ulem}
\usepackage{psfrag}
\usepackage{url}
\usepackage{babel}
\usepackage{authblk}
\usepackage[linesnumbered,ruled]{algorithm2e}
\usepackage[left=54pt, right=54pt, top=54pt, bottom=54pt]{geometry}
\usepackage{xspace}
\usepackage{multicol}
\usepackage{booktabs} 
\usepackage{multirow}
\usepackage[x11names,dvipsnames,table]{xcolor}

\usepackage{subfigure}

\usepackage{stackrel}
\usepackage{extarrows}

\usepackage{setspace}

\newcommand{\Roy}{\textcolor{blue}}
\newcommand{\confgraph}{\Gamma_t^{i}}
\newcommand{\confgraphtone}{\Gamma_{t_1}^{i,\ell}}
\newcommand{\confgraphttwo}{\Gamma_{t_2}^{i,\ell}}
\newcommand{\defective}{{\text{relaxed}}}
\newcommand{\Defective}{{\text{Relaxed}}}

\newcommand{\party}{{\text{party}}}

\newcommand{\opertype}[1]{\begin{minipage}{29mm}\centering\vspace{1mm} #1\vspace{1mm}\end{m
inipage}}

\newtheorem{theorem}{Theorem}
\newtheorem{lemma}{Lemma}

\newtheorem{proposition}{Proposition}
\newtheorem{observation}{Observation}
\newtheorem{example}{Example}
\newtheorem{definition}{Definition}
\newtheorem{remark}{Remark}

\DeclareMathOperator*{\argmax}{arg\,max}

\include{defns}  
\allowdisplaybreaks

\begin{document}
\title{On Converse Results for Secure Index Coding}

\author{Yucheng Liu$^{\dag}$\thanks{This work was supported by the ARC Discovery Scheme DP190100770, and the ARC Future Fellowship FT190100429.}, Lawrence Ong$^{\dag}$, Parastoo Sadeghi$^{*}$, Neda Aboutorab$^{*}$, and Arman Sharififar$^{*}$ \\\vspace{-0mm}
$^{\dag}$The University of Newcastle, Australia (emails: \{yucheng.liu, \hspace{0mm}lawrence.ong,\}@newcastle.edu.au)\\
$^{*}$University of New South Wales, Australia (emails: \{p.sadeghi,n.aboutorab,a.sharififar\}@unsw.edu.au)\vspace{0mm}
}


\maketitle

\input{abstract_allerton_2019.tex}






\input{main_results.tex}

\input{appendices.tex}

\bibliographystyle{IEEEtran}
\bibliography{references} 

\end{document}

%% file: defns.tex
\usepackage{xspace}
\usepackage{bbm}
\input{mathlig}

\newcommand{\muspace}{\mspace{1mu}}

\DeclareRobustCommand{\scond}{\mathchoice{\muspace\vert\muspace}{\vert}{\vert}{\vert}}
\mathlig{|}{\scond}

\DeclareRobustCommand{\discint}{\mathchoice{\mspace{-1.5mu}:\mspace{-1.5mu}}{\mspace{-1.5mu}:\mspace{-1.5mu}}{:}{:}}
\mathlig{::}{\discint}

\newcommand{\Ec}{\mathcal{E}}

\newcommand{\Hc}{\mathcal{H}}
\newcommand{\Ic}{\mathcal{I}}

\newcommand{\Lc}{\mathcal{L}}

\newcommand{\Nc}{\mathcal{N}}

\newcommand{\Sc}{\mathcal{S}}
\newcommand{\Tc}{\mathcal{T}}

\newcommand{\Vc}{\mathcal{V}}

\newcommand{\Xc}{\mathcal{X}}
\newcommand{\Yc}{\mathcal{Y}}
\newcommand{\Zc}{\mathcal{Z}}

\newcommand{\Ccal}{\mathcal{C}}

\newcommand{\Gcal}{\mathcal{G}}

\newcommand{\Pcal}{\mathcal{P}}
\newcommand{\Qcal}{\mathcal{Q}}

\newcommand{\Xcal}{\mathcal{X}}

\newcommand{\Zcal}{\mathcal{Z}}

\newcommand{\Cr}{\mathscr{C}}

\newcommand{\Xv}{{X}}

\newcommand{\Rv}{{\bf R}}

\newcommand{\tv}{{\bf t}}
\newcommand{\xv}{{x}}

\newcommand{\Xh}{{\hat{X}}}

\newcommand{\xh}{{\hat{x}}}

\newcommand{\It}{{\tilde{I}}}

\newcommand{\Yt}{{\tilde{Y}}}

\newcommand{\xt}{{\tilde{x}}}

\newcommand{\chit}{{\tilde{\chi}}}

\def\e{\epsilon}

\let\P\relax
\DeclareMathOperator\P{\textsf{P}}

\def\textiid{i.i.d.\@\xspace}
\newcommand\iid{\ifmmode\text{ i.i.d. } \else \textiid \fi}

\def\clap#1{\hbox to 0pt{\hss#1\hss}}
\def\mathclap{\mathpalette\mathclapinternal}
\def\mathclapinternal#1#2{%
  \clap{$\mathsurround=0pt#1{#2}$}}

\let\oldstackrel\stackrel
\renewcommand{\stackrel}[2]{\oldstackrel{\mathclap{#1}}{#2}}

\newcommand{\ntouch}[1]{\overline{{#1}}}

\newcommand{\Xhv}{\bf{\hat{X}}}

%% file: abstract_allerton_2019.tex
\begin{abstract}

In this work, we study the secure index coding problem where there are security constraints on both legitimate receivers and eavesdroppers. We develop two performance bounds (i.e., converse results) on the symmetric secure capacity. 
The first one is an extended version of the basic acyclic chain bound (Liu and Sadeghi, 2019) that takes security constraints into account. The second converse result is a novel information-theoretic lower bound on the symmetric secure capacity, which is interesting as all the existing converse results in the literature for secure index coding give upper bounds on the capacity.

\end{abstract} 


%% file: main_results.tex

\section{Introduction}\label{sec:intro}

Index coding \cite{Birk--Kol1998,bar2011index} studies the communication problem where a server broadcasts messages to multiple receivers with side information via a noiseless channel. 
In the classic \emph{multiple-unicast} setup of index coding, the server aims at delivering each message to one unique receiver. That is, there is a bijective function between the set of messages and the set of receivers, and each receiver needs to be able to decode its corresponding message based on the broadcast codeword and its own side information. 
Roughly speaking, the objective of index coding is to maximize the amount of information can be delivered to the receivers per channel use. 
For an excellent survey on index coding, see \cite{arbabjolfaei2018fundamentals}. 

The \emph{security} aspect of index coding has been studied in \cite{dau2012security,ong2016secure,liu:vellambi:kim:sadeghi:itw18,ong2021code}, where in addition to the legitimate receivers there is an eavesdropper, and the server must simultaneously satisfy the receivers' decoding requirements and protect the messages from being decoded by the adversary. 
%
An alternative setup where there is no external eavesdropper and the security constraints are against the receivers themselves, has been briefly discussed in \cite{dau2012security} and later studied in \cite{private:index:coding:arxiv,liu2020secure}. 

\begin{figure}[ht]
\begin{center}
\includegraphics[scale=0.2]{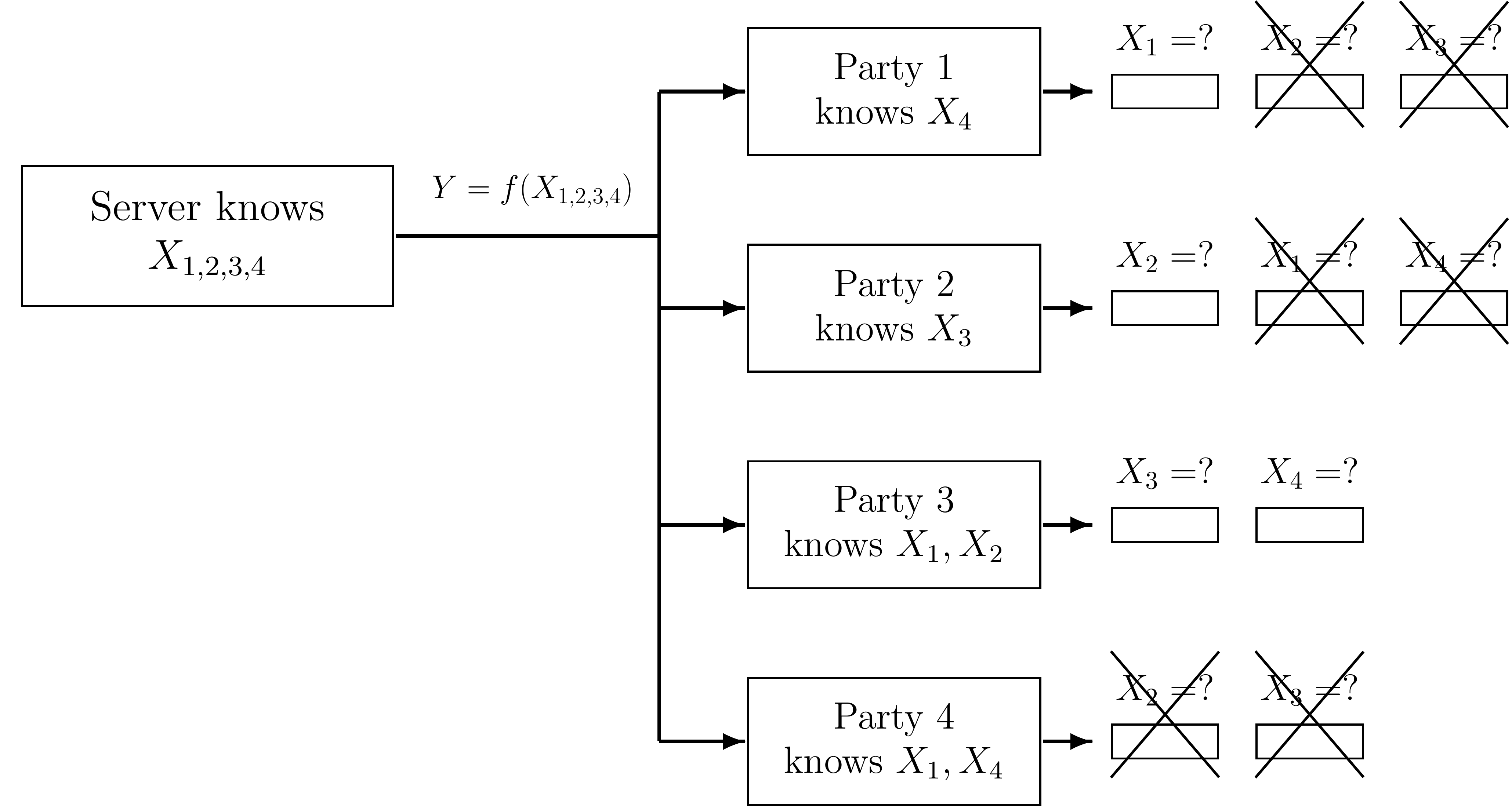}
\caption{A toy example where there are $4$-messages, $X_{1,2,3,4}=(X_1,X_2,X_3,X_4)$, stored at a server. 
The server broadcasts a codeword $Y$, which is function of $X_{1,2,3,4}$, and there are $4$ parties who observes the broadcast codeword, including $3$ receivers, and one eavesdropper. 
For each party, the server needs to ensure that it can decode certain message based on the broadcast codeword $Y$ and its own side information, and at the same time, cannot learn any individual message from a certain list. 
\vspace{-6mm}
}
\label{fig:secure:index:coding}
\end{center}
\end{figure}

In this work, we consider the secure index coding problem in a generic \emph{multiple-groupcast} setup that takes security constraints both against legitimate receivers and against the eavesdroppers into account. 
We consider a system where there are a number of messages and an arbitrary number of (legitimate) receivers and eavesdroppers. 
Each party observing the broadcasted codeword, whether a receiver or an eavesdropper, knows a fixed (but arbitrary) set of messages as side information, and the server tries to deliver a fixed (but arbitrary) set of its unknown messages to it. 
The server also needs to ensure that each party cannot learn any individual message from a certain prohibited message list for the party. 
In such a way, an eavesdropper can be simply seen as a special receiver for whom the server tries to deliver none of its unknown messages. 
See Figure \ref{fig:secure:index:coding} for a toy example of the considered system model.

The contributions and organization of the paper are as follows. 
We formally define our secure index coding problem in Section \ref{sec:model:setup}. Then in Section \ref{sec:model:preliminaries}, we extend several existing converse results from multiple-unicast setting to the multiple-groupcast setting considered here. 
Based upon those existing results, in Sections \ref{sec:sbac}, we develop our first new converse bound, namely, the secure basic acyclic chain lower bound on the symmetric \emph{secure} capacity. 
In Section \ref{sec:converse:upper}, we introduce our second main contribution, a novel information-theoretic lower bound on the symmetric {secure} capacity.

For non-negative integers $a$ and $b$, $[a]\doteq \{1,2,\cdots,a\}$, and $[a:b]\doteq \{a,a+1,\cdots,b\}$. If $a>b$, $[a:b]=\emptyset$. 
For any discrete random variable $Z$ with probability distribution $P_Z$, we denote its alphabet by $\Zc$ with realizations $z\in \Zc$. 

\section{System Model and Preliminaries}  \label{sec:model}


\subsection{System Model and Problem Setup}  \label{sec:model:setup}

Assume that there are $n$ messages, $x_i \in \{0,1\}^{t_i}, i \in [n]$, where $t_i\in {\mathbb Z}^{0+}$ is the length of binary message $x_i$. 
For brevity, when we say message $i$, we mean message $x_i$. 
Let $X_i$ be the random variable corresponding to $x_i$. We assume that $X_1, \ldots, X_n$ are independent and uniformly distributed. 
For any $S\subseteq [n]$, set $S^c \doteq [n]\setminus S$, $\xv_S \doteq (x_i,i\in S)$, and $\Xv_S \doteq (X_i,i\in S)$. 
By convention, $\xv_{\emptyset} = \Xv_{\emptyset} = \emptyset$. 
Set $N\doteq 2^{[n]}$ denotes the power set of the message set $[n]$. 

A server contains all messages $\xv_{[n]}$. 
Let $y$ be the output of the server, which is a deterministic function of $x_{[n]}$. 
The codeword $y$ is transmitted from the server to all legitimate receivers through a noiseless broadcast channel of normalized unit capacity, and also eavesdropped by a group of eavesdroppers. 
Let $Y$ be the random variable corresponding to $y$. 

There are in total $m$ receivers and eavesdroppers, indexed from $1$ to $m$. For any $i\in [m]$, the corresponding party (whether a receiver or an eavesdropper) knows some messages $x_{A_i}$ as side information for some $A_i\subseteq [n]$. 
The server needs to ensure that party $i$, based on the transmitted codeword $y$ and its side information $x_{A_i}$, can decode its (legitimately) requested messages $x_{W_i}$ for some set $W_i\subseteq A_i^c$. 
If party $i$ is a legitimate receiver, $W_i$ is a non-empty subset of $A_i^c$, and if {\party} $i$ is an eavesdropper, $W_i$ is simply an empty set. 
The set of indices of \emph{interfering messages} at party $i$ is denoted by the set $B_i\doteq (A_i\cup W_i)^c$.

The server also needs to satisfy certain security constraints against the receivers and eavesdroppers. 
That is, for each {\party} $i\in [m]$, there is a set of messages $P_i\subseteq B_i$, which the party is prohibited from learning. 
More specifically, party $i$ should not be able to infer any information about each individual message $j\in P_i$ given the side information $\xv_{A_i}$ and the received codeword $y$.\footnote{In \cite{dau2012security,ong2016secure,liu:vellambi:kim:sadeghi:itw18}, the eavesdropper is not allowed to learn \emph{any} message apart from its own side information. That is, $P_i=B_i$ for eavesdropper $i$. Here we consider a slightly more general setup that allows any $P_i\subseteq B_i$.} 


More formally, 
a $(\tv,M) = ((t_i, i \in [n]),M)$ {\em secure index code} is defined by
\begin{itemize}
\item A deterministic encoder at the server, $$\phi: \prod_{i \in [n]} \{0,1\}^{t_i} \to \{1,2,\ldots,M\},$$which maps the messages $\xv_{[n]}$ to a codeword index $y$; 
\item $m$ deterministic decoders, one for each {\party} $i\in [m]$, such that $$\psi_i:  \{1,2,\ldots,M\} \times \prod_{k \in A_i} \{0,1\}^{t_k} \to \{0,1\}^{t_i}$$ maps the received codeword $y$ and the side information $\xv_{A_i}$ back to $x_{W_i}$. 
\end{itemize}

We consider only surjective encoding functions $\phi$, meaning that each $y\in \Yc=[M]$ must be mapped from at least one $x_{[n]}$. This is because codewords not mapped from any message tuple realization will never be used and hence can be discarded.  
We say a rate tuple $\mathbf{R} = (R_i, i \in [n])$ is \emph{securely} achievable if 
there exists a $(\tv,M)$ index code satisfying 
\begin{alignat}{2}
&\text{\bf Rate:}                   && \mathbin{R_i=\frac{t_i}{\log M}, \qquad  \forall i \in [n],}      \label{con:model:rate}    \\
&\text{\bf Message:}             && H(\Xv_S|\Xv_{S'})=H(\Xv_S)=\sum_{i\in S}t_i,     \nonumber     \\
& &&  \qquad \qquad \forall S,S'\subseteq [n],S\cap S'=\emptyset,  \label{con:model:message} \\
&\text{\bf Codeword:}            && H(Y)\le \log M,    \label{con:model:codeword}    \\
&\text{\bf Encoding:}             && H(Y|\Xv_{[n]})=0,   \label{con:model:encoding}    \\
&\text{\bf Decoding:}   \quad  && H(X_{W_i}|Y,\Xv_{A_i})=0,    \qquad  \forall i\in [m],  \label{con:model:decoding}    \\
&\text{\bf Security:}              && I(X_j;Y|\Xv_{A_i})=0,  \qquad   \forall j\in P_i,i\in [m],     \label{con:model:security}  
\end{alignat}
where \eqref{con:model:rate} is the definition of $R_i,i\in [n]$, 
\eqref{con:model:message} follows from the assumption that the messages are independent and uniformly distributed, 
\eqref{con:model:codeword} is due to the size of the codeword alphabet being $M$, 
\eqref{con:model:encoding} follows from that $y$ is a deterministic function of $\xv_{[n]}$, 
\eqref{con:model:decoding} is stipulated by the zero-error decoding requirement at the legitimate receivers (note that for eavesdropper $i$, \eqref{con:model:decoding} becomes trivial as $W_i=\emptyset$), 
and \eqref{con:model:security} is stipulated by the security constraints on the receivers and eavesdroppers. 

\begin{remark}  \label{rmk:R:does:not:imply:smaller:R'}
In the trivial case where $t_1=t_2=\cdots=t_n=0$, we have $M=1$ as we only consider surjective deterministic encoding functions and we define the rate $R_i$ to be zero for any $i\in [n]$. 
For any valid index code with $t_i>0$ for some $i\in [n]$, we always have $M\ge 2$ as $M=1$ violates the decoding requirement in \eqref{con:model:decoding}. 
%
Also note that according to our definition of secure index codes, 
considering any rates $R> R'>0$, 
rate $R$ being securely achievable \emph{does not} imply that $R'$ is also securely achievable. 
\end{remark}


The secure capacity region $\Cr$ is 
the set of all securely achievable rate tuples $\Rv$. The symmetric secure capacity is defined as 
\begin{align} \label{eq:def:sym:C}
C\doteq \max \{ R:(R,\cdots,R) \in \Cr \}.
\end{align}

The decoding requirements and security constraints in \eqref{con:model:decoding}-\eqref{con:model:security} may be conflicting for some secure index coding problems, resulting in zero symmetric secure capacity $C=0$. We call such a problem infeasible. For more details, see \cite{liu2020secure}. 
For the rest of the paper, unless otherwise stated, we consider only feasible secure index coding problems for which $C>0$.


Any secure index coding problem can be compactly represented by a sequence $(W_i|A_i,P_i),i\in [m]$, specifying the requested messages, side information availability, and security constraints at the receivers and eavesdroppers. 
For example, the toy example in Figure \ref{fig:secure:index:coding} can be written as 
\begin{align*}
\begin{array}{cc}
(\{1\}|\{4\},\{2,3\}), & (\{2\}|\{3\},\{1,4\}), \\
(\{3,4\}|\{1,2\},\emptyset), & (\emptyset|\{1,4\},\{2,3\}).
\end{array}
\end{align*}


\subsection{Extension of Some Existing Results}  \label{sec:model:preliminaries}

Several useful performance bounds have been established in \cite{liu2020secure} under the multiple-unicast setup. 
Those results can be readily extended to the more generalized multiple-groupcast setting considered in this paper, where there can be more than one receiver demanding the same message. 
We present these extended performance bounds as follows. 
\begin{proposition}[Secure polymatoridal (S-PM) bound, \cite{liu2020secure}]   \label{thm:spm}
Consider a given secure index coding problem $(W_i|A_i,P_i)$, $i\in [m]$ with any valid $(\tv,r)$ secure index code. 
If a rate tuple $\Rv=(R_i,i\in [n])$ is securely achievable, it must satisfy
\begin{align}
&\sum_{i\in W}R_i=g(B\cup W)-g(B)=g(W),  \nonumber  \\
&\qquad \qquad \forall W\subseteq W_i,B\subseteq B_i\cup W_i\setminus W,i\in [m],\label{eq:Ri}
\end{align}
for at least one set function $g(S),S\in N=2^{[n]}$, such that
\begin{align}
&g(\emptyset)=0,  \label{eq:empty}  \\
&g([n])\le 1,  \label{eq:less:than:1}  \\
&g(S)\le g(S'),    \qquad    \text{if $S\subseteq S'$},  \label{eq:monotone}  \\
&g(S\cap S')+g(S\cup S')\le g(S)+g(S'),    \label{eq:submod} \\
&g(B_i)=g(B_i\setminus \{j\}),    \qquad    \forall j\in P_i,i\in [m]. \label{eq:securitynew} 
\end{align}
\end{proposition}

The above proposition can be proved by showing that the following set function 
\begin{align}
g(S)\doteq (\log M)^{-1}H(Y|\Xv_{S^c}),  \quad \forall S\subseteq [n],    \label{eq:g:def}
\end{align}
satisfies \eqref{eq:Ri}-\eqref{eq:securitynew} according to the system constraints \eqref{con:model:rate}-\eqref{con:model:security}. 
For more details, see \cite{liu2020secure} and \cite[Section 5.2]{arbabjolfaei2018fundamentals}.

The security property \eqref{eq:securitynew} is the major difference between the S-PM bound and the original PM bound \cite{blasiak2011lexicographic} for the non-secure index coding, which can be seen as a direct consequence of the security constraint \eqref{con:model:security}. 
It enforces the value of any set function $g$ satisfying \eqref{eq:securitynew} to be equal for certain arguments. Moreover, combining properties \eqref{eq:monotone} and \eqref{eq:submod} of $g$ with \eqref{eq:securitynew} may result in $g$ to be equal for even more arguments.
Based on such ideas, a partition on set $N=2^{[n]}$, namely the $\rm g$-partition, denoted by $\Nc=\{ N_1,N_2,\cdots,N_{\gamma} \}$, has been defined \cite{liu2020secure} as follows. 

\begin{definition}[{\rm g}-partition,\cite{liu2020secure}]  \label{def:gpartition}
Given a secure index coding problem $(W_i|A_i,P_i),i\in [m]$, its $g$-partition $\Nc$ can be constructed using the following steps:
\begin{enumerate}
\item \label{step:gpartition:1} For any receiver $i\in [m]$ whose $P_i$ is non-empty, for any $T\subseteq B_i^c$, form $N(i,T)\in \Nc$ as 
\begin{align*}
N(i,T)=\{ T\cup B_i\setminus \{j\}:j\in P_i \} \cup \{ T\cup B_i \}.
\end{align*} 
Note that $N(i,T)$ is a subset of $N$. 
All elements $S\in N$ that are not in any subset $N(i,T)$ are placed in $N_0$, i.e., 
\begin{align*}
N_0=N\setminus (\cup_{T\subseteq B_i^c,i\in [m]:P_i\neq \emptyset}N(i,T)). 
\end{align*}
\item As long as there exist two subsets $N(i,T),N(i',T')$ such that $N(i,T)\cap N(i',T')\neq \emptyset$, 
we merge these two subsets into one new subset. 
We keep merging overlapping subsets until all subsets in $\Nc$ are disjoint. 
Then we use $\gamma$ to denote the number of remaining subsets and index the elements of $\Nc$ as $N_1,N_2,\cdots,N_{\gamma}$ in an arbitrary order, except for $N_{\gamma}=N_0$. 
\label{step:gpartition:2}
\end{enumerate}
\end{definition}


For a given secure index coding problem $(W_i|A_i,P_i),i\in [m]$, due to the construction in Step \ref{step:gpartition:2} in the above definition, its $\rm g$-partition $\Nc$ is unique. We call any component within the $\rm g$-partition except for the last one a $\rm g$-subset. 
And for any message subsets $S,S'\subseteq [n]$, we write $S - S'$ iff they belong to the same $\rm g$-subset. 
Then we have the following lemma. 
\begin{lemma}[\cite{liu2020secure}]  \label{lem:g:subset}
Consider any set function $g$ satisfying \eqref{eq:Ri}-\eqref{eq:securitynew}. 
For any $S,S'\subseteq [n]$ such that $S-S'$, we have
\begin{align}
g(S)=g(S').
\end{align}
\end{lemma}

%


The maximum acyclic induced subgraph (MAIS) bound \cite{bar2011index} is a widely used performance bound for the non-secure index coding problem under multiple-unicast setup. 
It can be readily extended to the multiple-groupcast setting as follows. 
\begin{definition}[Acyclic set]  \label{def:acyclic:set}
Consider any message subset $K=\{i_1,i_2,\ldots,i_k\}\subseteq [n]$ and another message subset $S\subseteq K^c$. If they satisfy that for any $\ell\in [k]$, there exists some receiver $j_{\ell}\in [m]$ such that $i_{\ell}\in W_{j_{\ell}}$ and $S\cup \{i_1,i_2,\ldots,i_{\ell}\}\subseteq B_{j_{\ell}}\cup W_{j_{\ell}}$, we call set $K$ an acyclic set with respect to set $S$. 
If some set $K$ is an acyclic set with respect to $\emptyset$, we simply say that $K$ is acyclic. 
\end{definition}

\begin{proposition}[Maximum acyclic induced subgraph (MAIS) bound, \cite{bar2011index}]
Consider a non-secure index coding problem $(W_i|A_i,\emptyset),i\in [m]$. 
The (non-secure) symmetric capacity $C$ is upper bounded by the MAIS upper bound $C_{\rm MAIS}$, which is defined as the reciprocal of the cardinality of the largest acyclic set $K\subseteq [n]$. 
That is, 
\begin{align}
C\le C_{\rm MAIS}\doteq \max_{K\subseteq [n]:\text{$K$ is acyclic}}\frac{1}{|K|}.
\end{align}
\end{proposition}

More generally, 
for any message subsets $S\subseteq S'\subseteq [n]$, we define the following notation, 
\begin{align}
h_{\rm MAIS}(S,S')\doteq \max_{K\subseteq S'\setminus S: \text{ $K$ is acyclic w.r.t. $S$}}|K|.  \label{eq:def:mais:K:S}
\end{align}

Then we simply have 
$C_{\rm MAIS}=\frac{1}{h_{\rm MAIS}(\emptyset,[n])}. $

\begin{proposition}[Secure maximum acyclic induced subgraph (S-MAIS) bound, \cite{liu2020secure}]   \label{prop:smais}
Consider a secure index coding problem $(W_i|A_i,P_i),i\in [m]$ with $\rm g$-partition $\Nc=\{ N_1,N_2,\cdots,N_{\gamma} \}$. 
Its symmetric secure capacity $C$ can be upper bounded by the S-MAIS upper bound $C_{\rm S-MAIS}$, which can be constructed as follows: 
\begin{enumerate}
\item \label{step:secure:mais:1} For any subset $N_k$, $k\in [\gamma]$, initialize $\rho_k$ as 
$$ \rho_k=\max_{S\in N_k}h_{\rm MAIS}(\emptyset,S). $$   
\item As long as there exist two $\rm g$-subsets $N_k,N_{\ell}$, $k\neq \ell \in [\gamma-1]$ such that there exist some sets $S'\in N_k$, $S\in N_{\ell}$ satisfying that $S\subseteq S'$, 
and that
$$h_{\rm MAIS}(S,S')+\rho_{\ell}>\rho_k,$$ 
update $\rho_k \leftarrow h_{\rm MAIS}(S,S')+\rho_{\ell}$.    \label{step:secure:mais:2}
\item If no such $N_k,N_{\ell}$ exist, set $C_{\rm S-MAIS}=1/(\max_{k\in [\gamma]}\rho_{k})$ and terminate the algorithm.     \label{step:secure:mais:3}
\end{enumerate}
\end{proposition}

%
%
%
%
%

\section{Secure Basic Acyclic Chain Bound}  \label{sec:sbac}

In this section, we present our first main result. 

Basic acyclic chain bound has been proposed \cite{yuchengliu2019isit} for non-secure index coding and shown to be strictly tighter than the MAIS bound \cite{bar2011index} and the internal conflict bound \cite{maleki2014index,jafar2014topological}. 
In the following, we extend the basic acyclic chain bound to capture the security constraints described in \eqref{con:model:security}.

Consider any secure index coding problem $(W_i|A_i,P_i)$, $i\in [m]$, any securely achievable rate $R$, and any set function $g$ satisfying \eqref{eq:Ri}-\eqref{eq:securitynew}.

\begin{lemma}  \label{lem:g:difference}
For any $S\subseteq S'\subseteq [n]$, we have 
\begin{align}
g(S')\ge g(S)+R\cdot h_{\rm MAIS}(S,S').  \label{eq:lem:g:difference:1}
\end{align}
Furthermore, for any $S\subseteq S'\subseteq [n]$ such that $S-S'$, we have 
\begin{align}
h_{\rm MAIS}(S,S')=0.  \label{eq:lem:g:difference:2}
\end{align}
\end{lemma}

\begin{IEEEproof}
Consider any set $K\subseteq S'\setminus S$ of cardinality $k=h_{\rm MAIS}(S,S')$ that is acyclic with respect to $S$. 
By Definition~\ref{def:acyclic:set}, $K$ can be denoted as $K=\{i_1,i_2,\ldots,i_k\}$ such that for any $\ell \in [k]$, there exists some $j_{\ell}\in [m]$ such that 
\begin{align}
i_{\ell}\in W_{j_{\ell}}, \text{ and } S\cup \{i_1,i_2,\ldots,i_{\ell}\}\subseteq B_{j_{\ell}}\cup W_{j_{\ell}}.  \label{eq:lem:g:difference:proof:acyclic:1}
\end{align}

Then we have 
\begin{align}
g(S')&=g(S,S'\setminus S)  \nonumber  \\
&\ge g(S,\{ i_1,i_2,\ldots,i_k \})  \nonumber  \\
&\ge g(S,\{ i_1,i_2,\ldots,i_{k-1} \})+R_{i_k}  \nonumber  \\
&\ge g(S,\{ i_1,i_2,\ldots,i_{k-2} \})+R_{i_{k-1}}+R_{i_k}  \nonumber  \\
&\ge \cdots  \nonumber  \\
&\ge g(S)+k\cdot R  \nonumber  \\
&=g(S)+h_{\rm MAIS}(S,S')\cdot R,  \nonumber
\end{align}
where the first inequality follows from \eqref{eq:monotone}, and the second to the last inequalities are due to \eqref{eq:lem:g:difference:proof:acyclic:1} and \eqref{eq:Ri}. 

Combining \eqref{eq:lem:g:difference:1} and Lemma \ref{lem:g:subset} leads to \eqref{eq:lem:g:difference:2}.
\end{IEEEproof}

For any $S,S'\subseteq [n]$, define
\begin{align}  \label{eq:def:h}
h(S,S')=\begin{cases}
&h_{\rm MAIS}(S,S'), \quad \text{if $S\subseteq S'$,}  \\
&0, \qquad \qquad \qquad \text{otherwise.}
\end{cases}
\end{align}

\begin{lemma}  \label{lem:g:chain}
Consider any set function $g$ satisfying \eqref{eq:Ri}-\eqref{eq:securitynew}. 
Consider any sequence of message subsets indexed as $L_1,L_2,\ldots,L_k\subseteq [n]$ such that for any $j\in [k-1]$, we have either $L_j-L_{j+1}$ or $L_j\subseteq L_{j+1}$. 
Then we have 
\begin{align}
g(L_k)\ge g(L_1)+R\cdot \sum_{\ell\in [k-1]} h(L_{\ell},L_{\ell+1}).
\end{align}
\end{lemma}

\begin{IEEEproof}
By Lemma \ref{lem:g:subset}, for any $j\in [k-1]$ such that $L_j-L_{j+1}$, we have 
\begin{align}
g(L_{j+1})=g(L_j)=g(L_j)+R\cdot h(L_j,L_{j+1}),  \label{eq:lem:g:chain:proof:1}
\end{align}
where the last equality is due to the fact that $h(L_j,L_{j+1})=0$ following the definition in \eqref{eq:def:h} as well as \eqref{eq:lem:g:difference:2} in Lemma \ref{lem:g:difference}. 

For any $j\in [k-1]$ such that $L_j\subseteq L_{j+1}$, Lemma \ref{lem:g:difference} gives
\begin{align}
g(L_{j+1})&\ge g(L_j)+R\cdot h(L_j,L_{j+1}).  \label{eq:lem:g:chain:proof:2}
\end{align}


Summing up \eqref{eq:lem:g:chain:proof:1} for all $j\in [k-1]$ such that $L_j-L_{j+1}$ and \eqref{eq:lem:g:chain:proof:2} for all $j\in [k-1]$ such that $L_j\subseteq L_{j+1}$ yields
\begin{align}
g(L_k)\ge g(L_1)+R\cdot \sum_{j\in [k-1]} h(L_j,L_{j+1}),
\end{align}
which completes the proof.
\end{IEEEproof}

For any set $L_1\subseteq [n]$, define 
\begin{align}
h(L_1)
&\doteq \max_{\substack{L_1,L_2,\ldots,L_k \subseteq [n]:\\ \text{for any $j\in [k-1]$,} \\ \text{$L_j-L_{j+1}$ or $L_j\subseteq L_{j+1}$}}} \sum_{j\in [k-1]} h(L_j,L_{j+1}).  \label{eq:def:h:secure}
\end{align}

Now we are ready to propose our new converse bound. 
\begin{theorem}[Secure Basic Acyclic Chain Bound]  \label{thm:sbac}
A set of messages $I=\{i_1,i_2,\ldots,i_m,i_{m+1}\}$ forms a secure basic acyclic chain, $Ch$, denoted as 
\begin{align*}
Ch: i_1 \xleftrightarrow{h(\{i_1,i_2\})} i_2 
     \xleftrightarrow{h(\{i_2,i_3\})} \cdots 
     \xleftrightarrow{h(\{i_m,i_{m+1}\})} i_{m+1},      
\end{align*}
if there exists some $S-\{i_1,i_{m+1}\}$ such that $h_{\rm MAIS}(\emptyset,S)=2$. 
We call the edge between $i_j$ and $i_{j+1}$ edge $j$, and $h(\{i_j,i_{j+1}\})$ the height of edge $j$. 
The symmetric secure capacity $C$ is upper bounded by the secure basic acyclic chain bound induced by the chain $Ch$ as 
\begin{align}
C\le C_{\rm S-BAC}(Ch)\doteq \frac{m}{1+m+\sum\limits_{j\in [m]} h(\{i_j,i_{j+1}\})}. \label{eq:thm:sbac:result}
\end{align}
Define the secure basic acyclic chain bound $C_{\rm S-BAC}$ as the maximum of $C_{\rm S-BAC}(Ch)$ over all possible chains, we have 
\begin{align}
C&\le C_{\rm S-BAC}\doteq \max_{Ch} C_{\rm S-BAC}(Ch). \label{eq:thm:sbac:result:C}
\end{align}
\end{theorem}

\begin{IEEEproof}
Consider any securely achievable symmetric rate $R$ and any set function $g$ satisfying \eqref{eq:Ri}-\eqref{eq:securitynew}. 
We have 
\begin{align}
g(\{i_1,i_{m+1}\})&\stackrel{(a)}{=}g(S)  \nonumber  \\
&\stackrel{(b)}{\ge} g(\emptyset)+R\cdot h_{\rm MAIS}(\emptyset,S)\stackrel{(c)}{=}2R,  \label{eq:thm:sbac:proof:terminals}
\end{align}
where (a) follows from Lemma \ref{lem:g:subset}, (b) follows from Lemma \ref{lem:g:difference}, and (c) follows from \eqref{eq:empty}. 

Consider any edge $j\in [m]$. 
Following Lemma \ref{lem:g:chain} and the definition in \eqref{eq:def:h:secure}, as well as \eqref{eq:less:than:1} and \eqref{eq:monotone}, we have
\begin{align}
1\ge g(\{ i_j,i_{j+1} \})+R\cdot h(\{ i_j,i_{j+1} \}).  \label{eq:thm:sbac:proof:1}
\end{align}
Summing up \eqref{eq:thm:sbac:proof:1} for all $j\in [m]$, we obtain
\begin{align}
m&\ge \sum_{j\in [m]} g(\{ i_j,i_{j+1} \})+R\cdot \sum_{j\in [m]} h(\{i_j,i_{j+1}\})  \\
&=g(\{i_1,i_2\})+g(\{i_2,i_3\})+g(\{i_3,i_4\})+\cdots  \nonumber  \\
&\quad +g(\{i_m,i_{m+1}\})+R\cdot \sum_{j\in [m]} h(\{i_j,i_{j+1}\})  \\
&\ge R+g(\{i_1,i_3\})+g(\{i_3,i_4\})+\cdots  \nonumber  \\
&\quad +g(\{i_m,i_{m+1}\})+R\cdot \sum_{j\in [m]} h(\{i_j,i_{j+1}\})  \\
&\ge 2R+g(\{i_1,i_4\})+\cdots  \nonumber  \\
&\quad +g(\{i_m,i_{m+1}\})+R\cdot \sum_{j\in [m]} h(\{i_j,i_{j+1}\})  \\
&\ge \cdots  \nonumber  \\
&\ge (m-1)R+g(\{i_1,i_{m+1}\})  \nonumber  \\
&\quad +R\cdot \sum_{j\in [m]} h(\{i_j,i_{j+1}\}),  \label{eq:thm:sbac:proof:horizontal:chain}
\end{align}
where the third to the last inequalities follow from \eqref{eq:Ri}, \eqref{eq:empty}, \eqref{eq:monotone}, and \eqref{eq:submod}. 
Combining \eqref{eq:thm:sbac:proof:terminals} and \eqref{eq:thm:sbac:proof:horizontal:chain}, together with the definition in \eqref{eq:def:sym:C}, leads to \eqref{eq:thm:sbac:result}. 
\end{IEEEproof}

The secure basic acyclic chain bound is always no looser than the secure MAIS bound. 

\begin{proposition}  \label{prop:sbac:better:than:smais}
We always have $C\le C_{\rm S-BAC}\le C_{\rm S-MAIS}$. 
\end{proposition}

The proof of Proposition \ref{prop:sbac:better:than:smais} is relegated to Appendix \ref{app:proof:prop:sbac:better:than:smais}. 


\begin{example}
Consider the following $10$-message $10$-party (i.e., $n=m=10$) secure index coding problem, which is denoted by $(W_i|B_i,P_i),i\in [m]$ rather than $(W_i|A_i,P_i),i\in [m]$ for brevity: 
\begin{align*}
\begin{array}{cc}
(\{1\}|\emptyset,\emptyset), & (\{2\}|\emptyset,\emptyset)  \\
(\{3\}|\{6\},\emptyset), & (\{4\}|\{2,3\},\emptyset), \\
(\{5\}|\{2,3,4\},\emptyset),  & (\{6\}|\emptyset,\emptyset), \\
(\{7\}|\{1,2,6\},\{2,6\}),  & (\{8\}|\{1,6\},\emptyset), \\
(\{9\}|\{1,6,8\},\emptyset), & (\{10\}|\{1,3,6\},\{1,6\}).
\end{array}
\end{align*}
For this problem, it can be verified that the secure MAIS bound gives $C\le C_{\rm S-MAIS}=\frac{1}{3}$. 
On the other hand, for any set function $g$ satisfying \eqref{eq:Ri}-\eqref{eq:securitynew}, since $B_{10}=\{1,3,6\}$ and $P_{10}=\{1,6\}$, by \eqref{eq:securitynew}, we have 
\begin{align*}
g(\{1,3\})=g(\{1,3,6\})=g(\{3,6\}),
\end{align*}
which means $\{1,3\}-\{3,6\}$. 
Also, it can be verified that $h_{\rm MAIS}(\emptyset,\{3,6\})=2$. 
Therefore, we have the following secure basic acyclic chain,
\begin{align*}
1 \xleftrightarrow{\enskip h(\{1,2\}) \enskip} 2
     \xleftrightarrow{\enskip h(\{2,3\}) \enskip} 3.      
\end{align*}
For $L_1=\{1,2\}$, set $L_2=\{1,6\}$, $L_3=\{1,6,8\}$, and $L_4=\{1,6,8,9\}$. 
Then we have $L_j-L_{j+1}$ or $L_j\subseteq L_{j+1}$ for any $j\in \{1,2,3\}$ (for $j=1$, we have $g(L_1)=g(\{1,2,6\})=g(L_2)$ according to \eqref{eq:securitynew} and $B_7=\{1,2,6\}$, $P_7=\{2,6\}$). 
Therefore, by \eqref{eq:def:h:secure}, 
\begin{align*}
h(\{1,2\})&\ge \sum_{j\in \{1,2,3\}}h(L_j,L_{j+1})=0+1+1=2.
\end{align*}
It can be verified that the above inequality is actually tight, and thus $h(\{1,2\})=2$. 
Similarly, for $L_1=\{2,3\}$, we can set $L_2=\{2,3,4\}$, and $L_3=\{2,3,4,5\}$. We have $L_j-L_{j+1}$ or $L_j\subseteq L_{j+1}$ for any $j\in \{1,2\}$. Therefore,
\begin{align*}
h(\{2,3\})&\ge h(L_1,L_2)+h(L_2,L_3)=1+1=2.
\end{align*}
It can be verified that the above inequality is tight, and thus $h(\{2,3\})=2$. 
Finally, by Theorem \ref{thm:sbac}, we have
\begin{align*}
C_{\rm S-BAC}=\frac{2}{1+2+h(\{1,2\})+h(\{2,3\})}=\frac{2}{7},
\end{align*}
which is strictly tighter then $C_{\rm S-MAIS}=\frac{1}{3}$. 
It can be verified that $R=\frac{2}{7}$ is an achievable symmetric rate by the secure fractional local partial clique covering scheme \cite[Theorem 1]{liu2020secure} and thus the converse result $C_{\rm S-BAC}=\frac{2}{7}$ is actually tight, establishing $C=\frac{2}{7}$ for this problem. 
\end{example}



\section{An information-theoretic upper bound on $C$ due to the security constraints}  \label{sec:converse:upper}

Consider any secure index coding problem $(W_i|A_i,P_i)$, $i\in [m]$. 
Based on its own side information, each party $j\in [m]$ may be able to decode more messages than it needs. For example, suppose there is some receiver $i$ who knows message $1$ as side information and requests to decode message $2$. 
There is another receiver $j$ who knows messages $1$ and $3$ as side information and requests to decode message $4$. 
For any valid secure index codes satisfying the decoding requirement at receiver $i$, it will also enable receiver $j$ to decode message $2$ (even though $x_2$ is not requested by receiver $j$) as whatever side information known by receiver $i$ is also known by receiver $j$. 
Furthermore, for any party $i\in [m]$, after decoding any message that is not in its side information $A_i$, it has effectively enlarged its side information set to $A_i\cup \{i\}$, and based on this enlarged side information set it may be able to decode even more messages. 

According to such ``decoding chain" rule (for more details, see for example \cite{liu2018information}), 
let $D_i\subseteq [n]\setminus A_i$ to denote the messages receiver $i$ can decode given \emph{any} valid secure index codes. 
Note that as we only consider feasible secure index coding problems, we must have $D_i\cap P_i=\emptyset$. 
Set $\ntouch{A_i}\doteq A_i\cup D_i$ to denote the collection of messages that will be known to party $i$ after decoding. 

We have the following converse upper bound on $C$. 

\begin{theorem}  \label{thm:converse:upper}
Consider any message subset $S=\{ i_1,i_2,\ldots,i_k \}\subseteq [n]$. If for any $\ell \in [k]$, there exists some party $j_{\ell}\in [m]$ satisfying $i_{\ell}\in P_{j_{\ell}}$ and $\{ i_1,i_2,\ldots,i_{\ell-1} \}\subseteq \ntouch{A_{j_{\ell}}}$, then for any securely achievable rate $R$, we have $$ R\ge \frac{1}{n-|S|}, $$
and subsequently, we have $C\ge \frac{1}{n-|S|}$. 
\end{theorem}

\begin{IEEEproof}
Consider any securely achievable rate $R$ achieved by some valid $(t,M)$ secure index code. 

Consider any $\ell\in [k]$. By \eqref{con:model:security}, we have
\begin{align}
0&=I(X_{i_{\ell}};Y|X_{\ntouch{A}_{j_{\ell}}})  \nonumber  \\
&=H(X_{i_{\ell}}|X_{\ntouch{A}_{j_{\ell}}})-H(X_{j_{\ell}}|Y,X_{\ntouch{A}_{j_{\ell}}})  \nonumber  \\
&=H(X_{i_{\ell}}|X_{\{i_1,\ldots,i_{\ell-1}\}})-H(X_{i_{\ell}}|Y,X_{\ntouch{A}_{j_{\ell}}})  \nonumber  \\
&\ge H(X_{i_{\ell}}|X_{\{i_1,\ldots,i_{\ell-1}\}})-H(X_{i_{\ell}}|Y,X_{\{i_1,\ldots,i_{\ell-1}\}})  \nonumber  \\
&=I(X_{i_{\ell}};Y|X_{\{i_1,\ldots,i_{\ell-1}\}}),  \label{eq:thm:converse:upper:proof:mutual:info:1}
\end{align}
where the inequality follows from the fact that conditioning cannot increase entropy and that $\{ i_1,i_2,\ldots,i_{\ell-1} \}\subseteq \ntouch{A_{j_{\ell}}}$. 
According to \eqref{eq:thm:converse:upper:proof:mutual:info:1} and the non-negativity of mutual information, we have
\begin{align}
I(X_{i_{\ell}};Y|X_{\{i_1,\ldots,i_{\ell-1}\}})=0.  \label{eq:thm:converse:upper:proof:mutual:info:2}
\end{align}

Summing up \eqref{eq:thm:converse:upper:proof:mutual:info:2} for all $\ell\in [k]$ yields
\begin{align}
I(X_S;Y)=0.
\end{align}
Hence, we have 
\begin{align}
|S|\cdot t=H(X_S)&=H(X_S)-I(X_S;Y)  \nonumber  \\
&=H(X_S|Y)=\sum_{y\in \Yc}P_Y(y)H(X_S|Y=y),  \nonumber
\end{align}
where the first equality follows from that the messages are independent and uniformly distributed as stated in \eqref{con:model:message}. 
The above equality, together with the fact that $H(X_S|Y=y)\le \log 2^{|S|\cdot t}=|S|\cdot t$ for any $y\in \Yc$, indicates that 
\begin{align}
H(X_S|Y=y)=|S|\cdot t, \quad \forall y\in \Yc. 
\end{align}
In other words, given any specific codeword $y$, all realizations of $X_S$ must be possible and equally likely (recall that we do not allow dummy $y$ that is mapped from no $x_{[n]}$). 
Consequently, for any $y$, there must be at least $|\Xc_S|=2^{|S|\cdot t}$ many message tuple realizations $x_{[n]}$ mapped to it. 
As we are considering deterministic encoding functions, the number of codewords $\Yc$ must be upper bounded as
\begin{align*}
|\Yc|\le \frac{|\Xc_{[n]}|}{|\Xc_S|}=2^{(n-|S|)\cdot t},
\end{align*}
and subsequently we have $$R=\frac{t}{\log M}\ge \frac{t}{\log 2^{(n-|S|)\cdot t}}=\frac{1}{n-|S|}, $$
which subsequently indicate that $C\ge \frac{1}{n-|S|}$. 
%
\end{IEEEproof}


\begin{remark}
Theorem \ref{thm:converse:upper} is, to the best of our knowledge, the first converse result that gives lower bound on the symmetric secure capacity for secure index coding. 
\end{remark}

%% file: appendices.tex
\appendices

\section{Proof of Proposition \ref{prop:sbac:better:than:smais}}  \label{app:proof:prop:sbac:better:than:smais}

\begin{IEEEproof}
%
%
%
%
%
The proof of the proposition for the case where $C_{\rm S-MAIS}=1/\rho_{\gamma}$ is relatively simple and omitted due to limited space. 
Consider the case where $C_{\rm S-MAIS}=1/\rho_{\ell}$ for some $\ell\in [\gamma]$. 
Without loss of generality, we assume that there exists some different $\rm g$-subset indices $\ell_1,\ell_2,\ell_3,\ldots,\ell_k\in [\gamma-1]$ such that $C_{\rm S-MAIS}=1/\rho_{\ell_1}$ and that there exists some message subsets $S'_{j+1}\subseteq S_j\subseteq [n]$ for every $j\in [k-1]$, satisfying that $S_1\in N_{\ell_1}$ and $S_k'\in N_{\ell_k}$, and that  
\begin{align}
S_j,S_j'\in N_{\ell_j}, \enskip \forall j\in [2:k-1],
\end{align}
and that 
\begin{align}
\rho_{\ell_j}=\rho_{\ell_{j+1}}+h_{\rm MAIS}(S_{j+1}',S_j), \quad \forall j\in [k-1].
\end{align}
Also without loss of generality assume that for the last index ${\ell_k}$, there exists some acyclic set $S=\{ i_1,i_2,\ldots,i_s \}\in N_{\ell_k}$ such that $\rho_{\ell_k}=|S|=s$ and that 
$\{ i_1,i_2,\ldots,i_{p-1} \}\subseteq B_{i_p}$ for any $p\in [s]$. 
Then, by Proposition \ref{prop:smais}, we have
\begin{align}
1/C_{\rm S-MAIS}=\rho_{\ell_1}&=\rho_{\ell_k}+\sum_{j\in [k-1]} h_{\rm MAIS}(S_{j+1}',S_j)  \nonumber  \\
&=s+\sum_{j\in [k-1]} h_{\rm MAIS}(S_{j+1}',S_j).
\end{align}

For any set $L_1\subseteq [n]$, define 
\begin{align}
&h(L_1)  \nonumber  \\
&\doteq \max_{\substack{L_1,L_2,\ldots,L_k \subseteq [n]:\\ \text{for any $j\in [k-1]$,} \\ \text{$L_j-L_{j+1}$ or $L_j\subseteq L_{j+1}$}}} \sum_{j\in [k-1]} h(L_j,L_{j+1}).  \label{eq:def:h:secure}
\end{align}

We can construct the following basic acyclic chain
\begin{align}
Ch: i_1 \xleftrightarrow{\enskip h(\{i_1,i_2\}) \enskip} i_2,
\end{align}
as $h_{\rm MAIS}(\{i_1,i_2\})=2$ by Definition \ref{def:acyclic:set} and \eqref{eq:def:mais:K:S}. 
Given the existence of the message subset sequence $$\{i_1,i_2\},S,S_k',S_{k-1},S_{k-1}',S_{k-2},S_{k-2}',\ldots,S_2,S_2',S_1,$$ by \eqref{eq:def:h:secure}, we have
\begin{align}
h(\{i_1,i_2\})
&\ge \sum_{j\in [k-1]}h_{\rm MAIS}(S_{j+1}',S_j)  \nonumber  \\
&\quad +\sum_{j\in [2:k-1]}h_{\rm MAIS}(S_j',S_j)+h_{\rm MAIS}(\{i_1,i_2\},S)  \nonumber  \\
&\stackrel{(a)}{=}\sum_{j\in [k-1]}h_{\rm MAIS}(S_{j+1}',S_j)+h_{\rm MAIS}(\{i_1,i_2\},S)  \nonumber  \\
&\stackrel{(b)}{=}\sum_{j\in [k-1]}h_{\rm MAIS}(S_{j+1}',S_j)+s-2,  \nonumber
\end{align}
where (a) follows from that $S_j'-S_j$ for any $j\in [2:k-1]$ and thus $h_{\rm MAIS}(S_j',S_j)=0$ by \eqref{eq:def:h} and \eqref{eq:lem:g:difference:2} in Lemma \ref{lem:g:difference}, and (b) follows from the fact that $S$ is an acyclic set. 

Therefore, we have
\begin{align*}
C_{\rm S-BAC}&\le \frac{1}{1+1+h(\{i_1,i_2\})}  \\
&\le \frac{1}{2+\sum_{j\in [k-1]}h_{\rm MAIS}(S_{j+1}',S_j)+s-2}  \\
&=C_{\rm S-MAIS},
\end{align*}
which completes the proof. 
\end{IEEEproof}